\documentclass[aps,reprint,superscriptaddress]{revtex4-2}

\usepackage[T1]{fontenc}
\usepackage[utf8]{inputenc}

\usepackage{amsmath,amssymb}
\usepackage{hyperref}

\usepackage{graphicx}
\graphicspath{{figures/}}

\usepackage{xcolor}
\definecolor{blue}{RGB}{0,0,255}
\definecolor{red }{RGB}{255,0,0}

\begin{document}

\interdisplaylinepenalty=2500

\title{Tunable Multi-Frequency-Band Generator and Detector of AC Signals Based on a Spin Hall Device with Two Antiferromagnetic Layers}

\author{Oleh Shtanko}
\email{olegshtanko2000@gmail.com}
\author{Volodymyr Prokopenko}
\affiliation{%
Educational and Scientific Institute of High Technologies, Taras Shevchenko National University of Kyiv,
Kyiv 01601, Ukraine
}%

\author{Oleksandr Prokopenko}
\email{oleksandr.prokopenko@gmail.com}
\affiliation{%
Educational and Scientific Institute of High Technologies, Taras Shevchenko National University of Kyiv, Kyiv 01601, Ukraine
}%
\affiliation{G. V. Kurdyumov Institute for Metal Physics, National Academy of Sciences of Ukraine, Kyiv 03142, Ukraine}

\begin{abstract}
Antiferromagnetic (AFM) spin Hall oscillators offer an electrical route to compact terahertz sources, while AFM spin Hall diodes can be used as a promising detectors of terahertz signals, however, their attainable frequency and the redout efficiency are constrained by the dynamics of a single AFM layer. 
We theoretically and numerically investigate a spin Hall multilayer junction Pt/AFM/Pt/AFM/Pt comprising two canted AFM layers, driven by independent drive DC currents flowing in the outer platinum (Pt) layers. 
Using a model based on the coupled Landau--Lifshitz--Gilbert--Slonczewski equations for four AFM sublattices and a phase-dependent two-interface spin Hall magnetoresistance model, we show that the device manifests itself as a DC current-driven oscillating spin Hall magnetoresistance. 
When an additional DC (AC) signal current is applied to the junction, the device generates an output AC (DC) voltage. 
Our model predicts current-driven, locked, co- and counter-rotational states of the net magnetization at N\'eel-vector frequencies, $f_{\rm T}$ and $f_{\rm B}$, in the top and bottom AFM layers, respectively ($f_{\rm T,B} \approx 0.5$~THz for a device with typical parameters). 
These states result in a multi-frequency response of the spin Hall magnetoresistance at the frequencies of $f_{\rm T} \pm f_{\rm B}$. 
The counter-rotation regime additionally activates the phase-sensitive mixing contribution, producing a fourth harmonic. 
In contrast, the co-rotation regime provides a difference-frequency channel on the order of 100~GHz, although mutual locking prevents this output from being tuned continuously toward zero. 
Our estimations show that the device matched to an external load can generate output AC power of 1.2~nW at the second harmonic and 0.8~pW at the fourth harmonic, and can produce a DC voltage of about $1$~mV for a 6~mA input signal current. 
The obtained results pave the way for the development and optimization of tunable AFM spintronic devices that can operate in a wide range of frequencies.
\end{abstract}


\maketitle

\section{Introduction}
\label{sec:introduction}

Antiferromagnetic (AFM) spintronics offers an intriguing opportunity to develop and improve various nanoelectronic devices \cite{Chen2024Emerging,Rimmler2025Noncollinear,Sakamoto2025Diode,Yao2025Transistors}.
The distinct advantages of AFM materials are crucial for the enhancement of high-speed electronics, successfully bridging a terahertz-frequency gap \cite{Cheng2025Antiferromagnet,Rene2025Terahertz}. 
Characterized by negligible macroscopic net magnetization, compensated antiferromagnets are inherently robust against parasitic external magnetic fields and emit no stray perturbations that could disrupt integrated circuitry \cite{Chen2024Emerging, Rimmler2025Noncollinear}.
Featuring N\'eel temperatures far exceeding room temperatures, the rich availability of AFM materials proves highly beneficial, enabling robust functionality without a dependence on extreme cryogenic cooling, large-scale experimental architectures, strong magnetic fields, or high DC voltages \cite{Cheng2025Antiferromagnet,Rene2025Terahertz,Ogawa2026Ultrafast}.

Exciting the ultrafast magnetic dynamics in AFM materials is possible through the application of the spin-transfer torque \cite{Lee2025SpinTorque}.
Specifically, driving a spin current through the device induces an antidamping torque that actively stimulates an ultra-fast magnetization dynamics.
As a result, when the spin current magnitude reaches a certain threshold, it triggers the oscillatory motion of the magnetic sublattices in an AFM layer \cite{Cheng2016Terahertz,Khymyn2017Josephson}.
This current-driven behavior provides a functional basis for creating and registering AC signals in the sub-terahertz and terahertz frequency bands.
The concept of exploiting these physical processes for current-driven magnetic dynamics in antiferromagnets was introduced in \cite{Gomonay2010SpinTransfer}, where the authors established the theoretical possibility of driving a steady precession of the AFM vector using an applied spin current.
This research has stimulated extensive subsequent studies proposing terahertz-frequency generators and detectors based on AFM spintronic nanostructures.

In a pioneering paper \cite{Cheng2016Terahertz}, researchers theoretically proposed the concept of a terahertz AFM spin Hall oscillator (SHO).
Employing the spin Hall geometry commonly used in spintronics \cite{Fischer2018SpinHall,Sinova2015SpinHall}, the authors investigated an insulating AFM/heavy metal heterostructure.
Within this setup, they showed that sufficiently large in-plane currents excite magnetization oscillations at terahertz frequencies.
Furthermore, the AC voltage output can be tuned by varying the applied DC current.
Building on this configuration, a subsequent study \cite{Khymyn2017Josephson} investigated the same spin Hall geometry in a bilayer comprising heavy metal platinum (Pt) layer and the biaxial AFM NiO layer.
Crucially, the research confirmed the structure's potential to act as a current-driven source of a terahertz-frequency electric field.
Using an extraction mechanism based on the inverse spin Hall effect in Pt, the researchers showed that the working principle of their room-temperature nano-oscillator resembles that of a Josephson junction \cite{Miyamoto2024WideBand,Karimi2024Bolometric}.
In this system, the easy-plane magnetic-anisotropy energy plays a role analogous to that of the Josephson energy. 
The further development of such systems involves studying their properties under the influence of pulse signals, where the device emulates the behavior of artificial neurons \cite{Khymyn2018Neuron,Sotnyk2026PNeuron} and neuron-based circuits \cite{Sulymenko2018Logic}.

As an alternative approach for terahertz-frequency power extraction, the utilization of canted AFM materials was introduced in \cite{Sulymenko2017Terahertz}.
Consequently, the rotating non-zero net magnetization of the AFM layer emits magnetic dipole radiation, which can then be captured by an electrodynamic system, such as a dielectric resonator.
For the resonator configurations analyzed in that work, the output power increases with frequency and can exceed 1~$\mu$W near 0.5~THz at experimentally attainable driving-current densities.
Moving beyond the conventional Pt/AFM spin Hall geometry, researchers in \cite{Sulymenko2018Tunnel} proposed an important alternative source of terahertz-frequency signal by exploring an AFM tunnel junction Pt/AFM/MgO/Pt.
By inserting an MgO layer between the AFM material and a second Pt electrode, the architecture forms a magnetic junction similar to the one explored in \cite{Park2011SpinValve}.
In such a device, stable current-driven dynamics is excited within the AFM material by the spin Hall effect \cite{Sinova2015SpinHall} and the spin-transfer torque, and the generated terahertz-frequency signal is successfully extracted through tunneling anisotropic magnetoresistance \cite{Xu2025SpinFlop}.
For optimized device geometry, the projected output power exceeds 1~$\mu$W at 0.5~THz, with an efficiency of at least 1\%.

Another promising application of antiferromagnet-based spin Hall heterostructures is their use as terahertz-frequency signal detectors and spectral analyzers. 
The first theoretical concept of an AFM/heavy metal spin Hall detector (SHD) was proposed in \cite{Khymyn2017SHD} and has been further developed in many subsequent papers (see \cite{Sakamoto2025Diode} and the review \cite{Prokopenko2024Book}).
While the considered device demonstrates the potential for detecting AC input signals in AFM nanostructures, its initial efficiency was rather limited.
A similar approach to that proposed for AFM SHOs was presented in \cite{Artemchuk2020ATJ}, in which an SHD based on an AFM tunnel junction was considered.
This SHD utilizes the tunneling anisotropic magnetoresistance effect and can generate a DC voltage of about 1~mV, which is sufficient for many applications.
Another prominent feature of the considered SHD is its capability to perform spectral analysis of terahertz-frequency input signals when the drive current applied to the device sweeps over a wide range \cite{Artemchuk2020ATJ}.

A key characteristic of all the discussed AFM devices is the intrinsic tunability of their generated terahertz-frequency signals: variations in the DC current density flowing through the Pt layer change the device operating frequency.
Therefore, the experimentally achievable values of the driving DC current density dictate the practical frequency limits for any signal source and signal detectors relying on an AFM layer and the spin Hall effect.

In this work, we investigate a dual-antiferromagnetic-layer (DAL) spin Hall oscillator and detector with a Pt/AFM/Pt/AFM/Pt architecture that demonstrates the potential to partially overcome the previously discussed operating frequency limitation. 
The independent excitation of the two AFM layers makes their relative phase an additional degree of freedom, while the shared central Pt electrode receives the mixing contribution and provides access to the multi-frequency difference signal generated by the rotating sublattices in the AFM layers. 
We consequently examine both the harmonic output associated with co- and counter-rotational dynamics and the range over which a nonzero difference-frequency response can be sustained before mutual synchronization locks the layers. 
Additionally, we estimate the output DC voltage generated in the DAL structure when an input harmonic AC signal is applied.

\begin{figure}
  \centering
  \includegraphics[width=\linewidth]{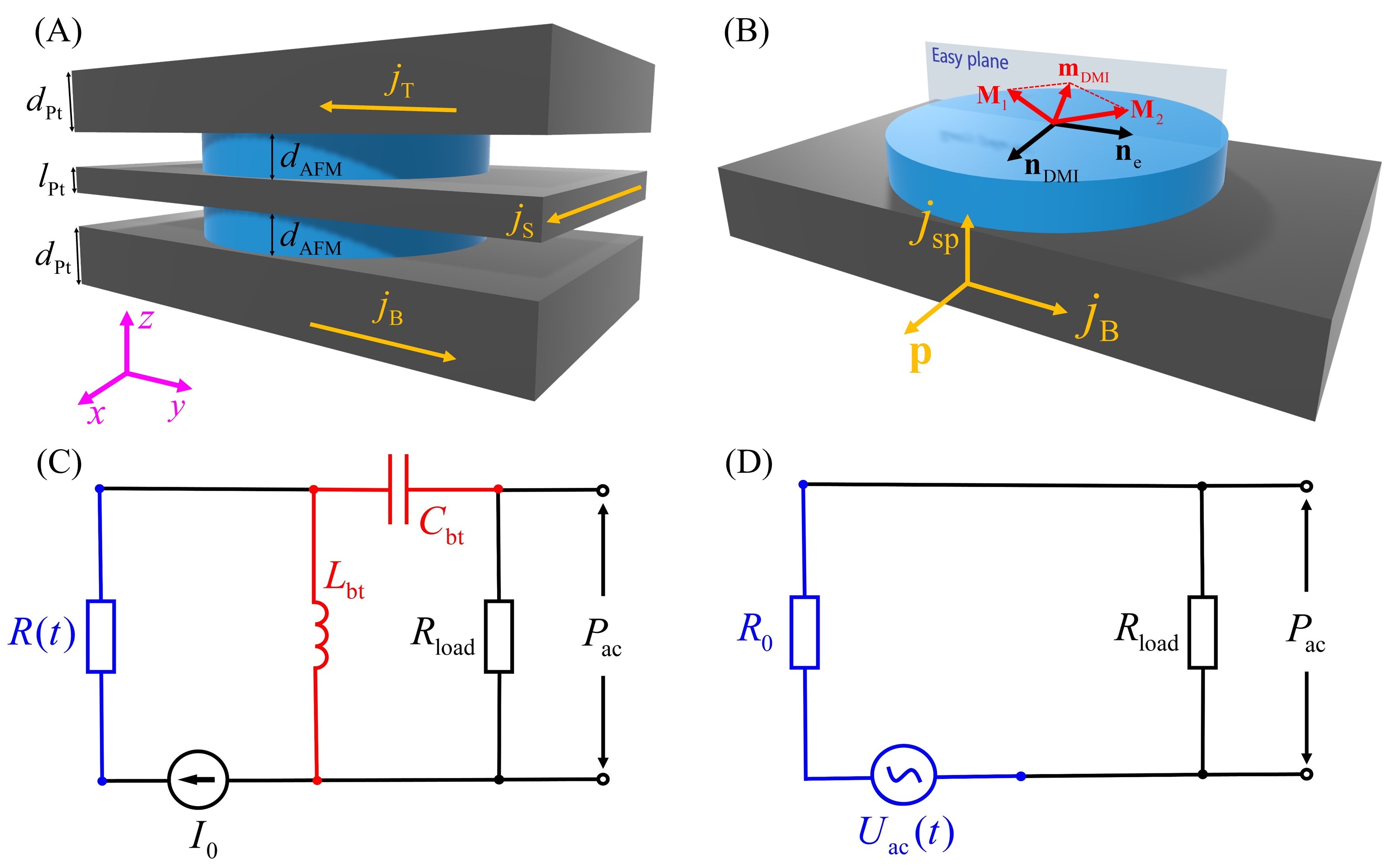}
  \caption{DAL spin Hall device architecture, local spin Hall geometry, and electrical readout. 
  (A) Pt/AFM$_{\mathrm{T}}$/central-Pt/AFM$_{\mathrm{B}}$/Pt stack. The outer Pt layers carry the signed drive-current densities $j_{\mathrm{T}}$ and $j_{\mathrm{B}}$; the central Pt layer carries the weak read signal current density $j_{\mathrm{S}}$. 
  (B) Canted AFM/Pt geometry: $\mathbf{M}_{1,2}$ are the sublattice magnetizations, $\mathbf{m}_{\mathrm{DMI}}=\mathbf{M}_{1}+\mathbf{M}_{2}$ is the DMI-induced net magnetization, and the shaded $yz$ plane is the easy plane. The outer-layer charge current produces transverse spin current $\mathbf{j}_{\mathrm{sp}}$ with polarization $\mathbf{p}=\hat{\mathbf{x}}$. 
  (C) The DC-biased circuit with an ideal bias tee used for the extraction of the central-Pt output signal: AC signal for the spin Hall oscillator and DC voltage for the spin Hall detector. (D) Small-signal Thevenin representation with DC resistance $R_0$, AC source $U_{\mathrm{ac}}(t)$, and load $R_{\mathrm{load}}$.}
  \label{fig:device-circuit}
\end{figure}

\section{Materials and Methods}\label{sec:methods}

The proposed dual-antiferromagnetic-layer (DAL) spin Hall device is a Pt/AFM$_{\mathrm{T}}$/Pt/AFM$_{\mathrm{B}}$/Pt heterostructure presented in Figure~\ref{fig:device-circuit}(A).
The top and bottom Pt layers are independent drive electrodes, while the shared central Pt layer is a weakly biased longitudinal readout (signal) electrode.
The outer Pt thickness is $d_{\mathrm{Pt}}=20$~nm, while the thickness of each canted AFM layer is $d_{\mathrm{AFM}}=5$~nm. These values are typical for AFM SHOs \cite{Khymyn2017Josephson,Sulymenko2017Terahertz}.
The thickness of the central Pt layer is $l_{\mathrm{Pt}}=12$~nm, corresponding to that used in the similar ferromagnetic structure studied in \cite{Prokopenko2013Hysteresis}.
In the macrospin model with dipolar coupling \cite{Prokopenko2013Hysteresis,Shtanko2024DualAFM}, the two coaxial AFM elements are assumed to have radius of $r=250$~nm.

The charge-current densities $j_{\mathrm{T}}$, $j_{\mathrm{S}}$, and $j_{\mathrm{B}}$ correspond to the top drive, signal (readout), and bottom drive currents, respectively. 
The outer (drive) currents generate damping-like spin Hall torques at their adjacent AFM/Pt interfaces.
The smaller current density, $j_{\mathrm{S}}$, is only used for the electrical readout of the generated signal (AC power for the oscillator and DC voltage for the detector).
The directions of the drive current densities, $j_{\mathrm{T}}$ and $j_{\mathrm{B}}$, are defined by the polarity of the applied DC voltages across the outer Pt electrodes. 
Figure~\ref{fig:device-circuit}(A) shows the drive currents flowing in opposite directions, which, as we will demonstrate later, results in a better high-frequency device performance. 
However, the two drive currents do not always have to be oppositely directed. 
When the top and bottom drive currents flow in the same direction, a low-frequency regime of the device operation can be achieved.

The magnetic configuration in the AFM layer of the DAL device is shown in Figure~\ref{fig:device-circuit}(B). 
Each AFM layer is made of hematite ($\alpha$-Fe${}_2$O${}_3$) and is modeled as a canted two-sublattice system with physical sublattice magnetizations $M_s\mathbf{m}_{a}^{(i)}$, where $M_s$ is the saturation magnetization, $a\in\{1,2\}$ labels the first and second sublattices, $i\in\{\mathrm{B},\mathrm{T}\}$ labels the bottom and top AFM layers, respectively, and $|\mathbf{m}_{a}^{(i)}|=1$.
The corresponding dimensionless N\'eel $\mathbf{l}^{(i)}$ and net $\boldsymbol{\mu}^{(i)}$ order parameters are
\begin{equation}
 \mathbf{l}^{(i)}=\frac{\mathbf{m}_{1}^{(i)}-\mathbf{m}_{2}^{(i)}}{2},
 \qquad
 \boldsymbol{\mu}^{(i)}=\frac{\mathbf{m}_{1}^{(i)}+\mathbf{m}_{2}^{(i)}}{2}.
 \label{eq:order-parameters}
\end{equation}
Here $\mathbf{l}^{(i)}$ is the order parameter used below, whereas $2M_s\boldsymbol{\mu}^{(i)}$ is the physical net magnetization shown by $\mathbf{m}_\mathrm{DMI}$ in Figure~\ref{fig:device-circuit}(B) as a result of the Dzyaloshinskii--Moriya interaction (DMI) \cite{Zhao2025InterlayerDMI,Yang2023FirstPrinciplesDMI}. 
The hard-axis, DMI, and outer-drive spin-polarization direction are the same, $\hat{\mathbf{x}}=\mathbf{n}_{\mathrm{h}}=\mathbf{n}_{\mathrm{DMI}}=\mathbf{p}$; $\hat{\mathbf{y}}=\mathbf{n}_{\mathrm{e}}$ is the in-plane anisotropy direction; and the easy plane is the $yz$ plane.

The magnetic dynamics are described by the coupled Landau--Lifshitz--Gilbert--Slonczewski (LLGS) equations \cite{Sulymenko2017Terahertz, Khymyn2017Josephson, Kim2016Ultrafast} for all four magnetic sublattices of the AFM layers,
\begin{eqnarray}
 \frac{d\mathbf{m}_{a}^{(i)}}{dt}&=&
 \gamma\mu_{0}\mathbf{H}_{a}^{(i)}\times\mathbf{m}_{a}^{(i)}+
 \alpha\,\mathbf{m}_{a}^{(i)}\times\frac{d\mathbf{m}_{a}^{(i)}}{dt}+
 \nonumber\\
 &&\tau^{(i)}\,\mathbf{m}_{a}^{(i)}\times
 \left(\mathbf{m}_{a}^{(i)}\times\hat{\mathbf{x}}\right)\,.
 \label{eq:llgs}
\end{eqnarray}
Here $\gamma$ is the gyromagnetic ratio modulus, $\mu_0$ is the vacuum permeability, and the Gilbert damping constant $\alpha=\alpha_0+\alpha_{\mathrm{sp}}$ includes contributions from intrinsic damping $\alpha_0$ and spin pumping \cite{Cheng2014SpinPumping,Hattori2024SpinPumping},
\begin{equation}
 \alpha_{\mathrm{sp}}=\frac{g^{\uparrow\downarrow}\gamma\hbar}{4\pi M_s d_{\mathrm{AFM}}}\,,
 \label{eq:spin-pumping}
\end{equation}
where $g^{\uparrow\downarrow}$ is the spin-mixing conductance at the Pt/AFM interface,
and $\hbar$ is the reduced Planck constant.
The signed drive density $j^{(i)}$ sets the spin-transfer torque amplitude of the top or bottom AFM layer \cite{Khymyn2017Josephson,Khymyn2018Neuron},
\begin{equation}
 \tau^{(i)}=\frac{j^{(i)} g^{\uparrow\downarrow}\theta_{\mathrm{SH}}e\gamma
 \lambda_{\mathrm{Pt}}\rho_{\mathrm{Pt}}}
 {2\pi M_s d_{\mathrm{AFM}}}
 \tanh\!\left(\frac{d_{\mathrm{Pt}}}{2\lambda_{\mathrm{Pt}}}\right)\,,
 \label{eq:torque-amplitude}
\end{equation}
where $\theta_\mathrm{SH}$, $\lambda_{\mathrm{Pt}}$, $\rho_{\mathrm{Pt}}$ and $e$ denote the spin Hall angle, spin-diffusion length, resistivity of the Pt layer, and the elementary electric charge, respectively.
For $\bar a=3-a$ denoting the opposite sublattice and $\bar i$ denoting the other AFM layer in the structure (e.g., $a = 1$ and $\bar a = 2$, $i = {\rm T}$ and $\bar i = {\rm B}$), the effective magnetic field is \cite{Sulymenko2017Terahertz,Prokopenko2013Hysteresis}
\begin{eqnarray}
 \mathbf{H}_{a}^{(i)}&=&{}-H_{\mathrm{ex}}\mathbf{m}_{\bar a}^{(i)}
 -H_{\mathrm{h}}\mathbf{n}_\mathrm{h}(\mathbf{n}_\mathrm{h}\!\cdot\!\mathbf{m}_{a}^{(i)})
 +H_{\mathrm{e}}\mathbf{n}_\mathrm{e}(\mathbf{n}_\mathrm{e}\!\cdot\!\mathbf{m}_{a}^{(i)})+
 \nonumber\\
 &&s_aH_{\mathrm{DMI}}\hat{\mathbf{x}}\times\mathbf{m}_{\bar a}^{(i)}
 -M_s\mathbf{N}_{\mathrm{cross}}\mathbf{m}_{a}^{(\bar i)} \, ,
 \nonumber\\
 &&\qquad s_a = \{s_1, s_2\} = \{-1, +1\} \, .
 \label{eq:effective-field}
\end{eqnarray}
These five terms in $\mathbf{H}_{a}^{(i)}$ represent the exchange field, hard-axis anisotropy field directed along the unit vector $\mathbf{n}_\mathrm{h}$, easy-plane anisotropy field directed along the unit vector $\mathbf{n}_\mathrm{e}$, DMI field, and the corresponding-sublattice mutual dipolar coupling field, respectively. 
For two coaxial circular canted AFM layers separated by the central nonmagnetic spacer we use the cross-demagnetization tensor $\mathbf{N}_{\mathrm{cross}}$ from \cite{Prokopenko2013Hysteresis,Dmytriiev2010CrossTensor},
\begin{eqnarray}
 \mathbf{N}_{\mathrm{cross}}&=&
 \begin{pmatrix}\rho&0&0\\0&\rho&0\\0&0&-2\rho\end{pmatrix}\,,
 \nonumber\\ 
 \rho&=&\frac{\ln 8-\frac{1}{2}}{2\pi}\frac{d_{\mathrm{AFM}}}{r}-
 \nonumber\\
 &&\frac{\mathcal{F}\left(\frac{l}{r}\right)-2\mathcal{F}\left(\frac{l+d_{\mathrm{AFM}}}{r}\right)+
 \mathcal{F}\left(\frac{l+2d_{\mathrm{AFM}}}{r}\right)}
 {4\pi\frac{d_{\mathrm{AFM}}}{r}}\,,
 \nonumber\\
 \qquad \mathcal{F}(\zeta)&=&\zeta^2\ln|\zeta| \, .
 \label{eq:cross-demag}
\end{eqnarray}

\begin{figure}
  \centering
  \includegraphics[width=\linewidth]{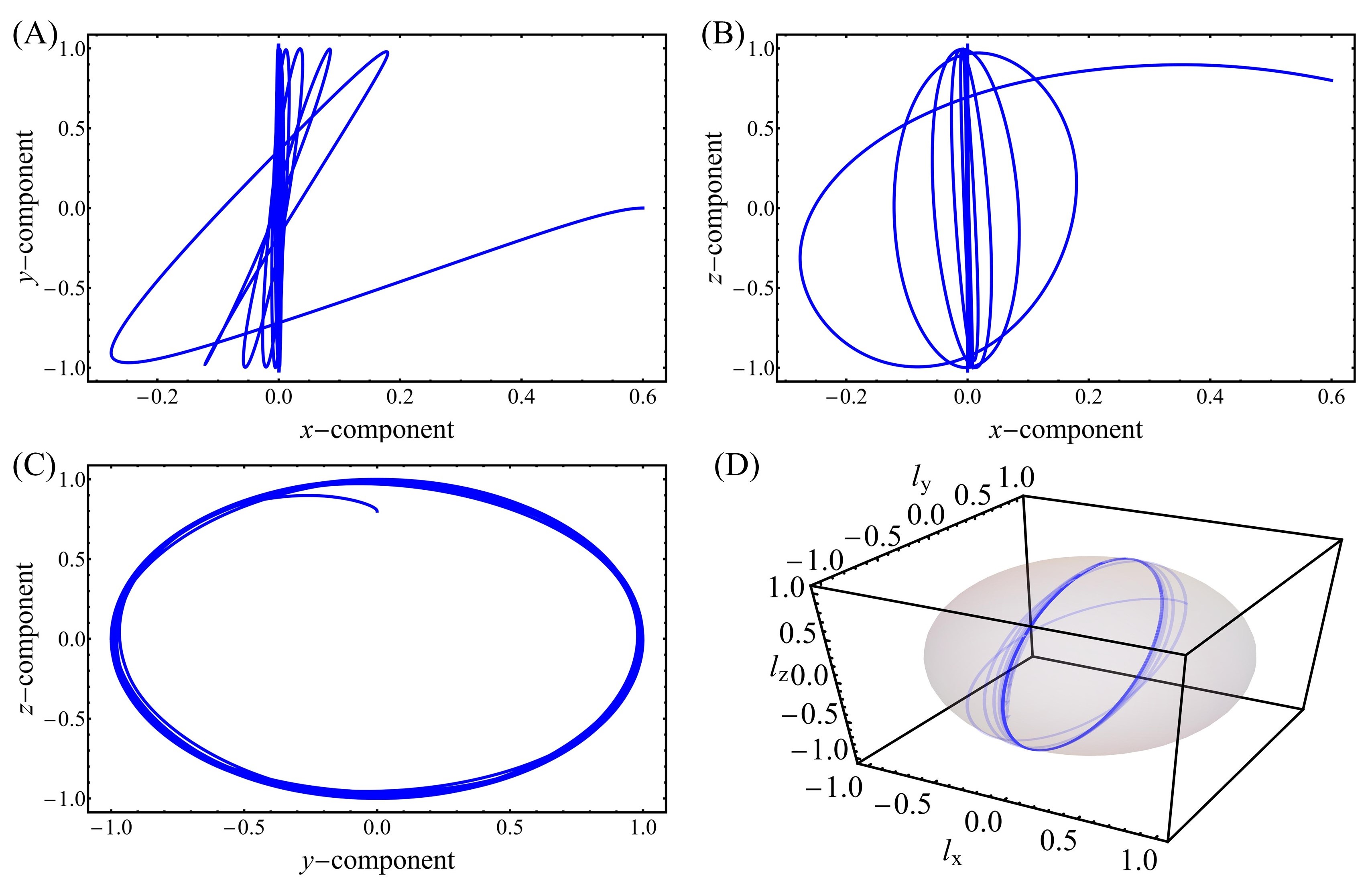}
  \caption{Trajectory of the bottom-layer N\'eel-vector ${\bf l}^{(B)}$ during 0--100~ps of DAL SHO dynamics. 
  (A) $l_x$--$l_y$ projection, 
  (B) $l_x$--$l_z$ projection, and 
  (C) $l_y$--$l_z$ projection of $\mathbf{l}^{(\mathrm{B})}(t)$. 
  (D) The corresponding normalized trajectory $\hat{\mathbf{l}}^{(\mathrm{B})}=\mathbf{l}^{(\mathrm{B})}/|\mathbf{l}^{(\mathrm{B})}|$ on the unit sphere. The initial non-closed segment is the transient from the prescribed antiparallel sublattice state; the recurrent part is the steady orbit.}
  \label{fig:neel-trajectory}
\end{figure}

The typical dynamics of the N\'eel vectors in a DAL SHO is illustrated in Figure~\ref{fig:neel-trajectory}.

The central Pt layer probes the unit N\'eel directions $\hat{\mathbf{n}}^{(i)}=\mathbf{l}^{(i)}/|\mathbf{l}^{(i)}|$. 
With read-signal-current spin-polarization $\hat{\mathbf{p}}_{\mathrm{S}}=\hat{\mathbf{y}}$, we define the following vector $\mathbf{v}_i=\hat{\mathbf{p}}_{\mathrm{S}}-(\hat{\mathbf{n}}^{(i)}\!\cdot\!\hat{\mathbf{p}}_{\mathrm{S}})\hat{\mathbf{n}}^{(i)}$, which determines the resistance in the two-interface spin Hall magnetoresistance (SHMR) model \cite{Chen2013SMR,Lebrun2019SMR},
\begin{equation}
 R(t)=R_{\mathrm{dc}}+a_{\mathrm{T}}|\mathbf{v}_{\mathrm{T}}|^2+
 a_{\mathrm{B}}|\mathbf{v}_{\mathrm{B}}|^2+
 \kappa\,\mathbf{v}_{\mathrm{T}}\!\cdot\!\mathbf{v}_{\mathrm{B}}.
 \label{eq:smr-coupled-resistance}
\end{equation}
Here $R_{\mathrm{dc}}$ is the device DC resistance, $a_{\mathrm{T,B}}$ and $\kappa$ are quantity given in the units of resistance, and $\kappa$ is the phenomenological shared-spin-accumulation coefficient. 
Lying in the easy plane, the N\'eel vector could be modeled as $\mathbf{l}^{(i)}=(0,\sin\phi_i,\cos\phi_i)$, yielding
\begin{eqnarray}
 R(t)&=&R_{\mathrm{off}}+R_{\mathrm{T}}\cos(2\phi_{\mathrm{T}})
 +R_{\mathrm{B}}\cos(2\phi_{\mathrm{B}})+
 \nonumber\\
 &&R_{\mathrm{mix}}\cos[2(\phi_{\mathrm{T}}-\phi_{\mathrm{B}})]\,,
 \label{eq:smr-phase-resistance}
\end{eqnarray}
where $R_{\mathrm{T}}=a_{\mathrm{T}}/2+\kappa/4$, $R_{\mathrm{B}}=a_{\mathrm{B}}/2+\kappa/4$, $R_{\mathrm{mix}}=\kappa/4$, and $R_{\mathrm{off}}=R_{\mathrm{dc}}+(a_{\mathrm{T}}+a_{\mathrm{B}})/2+\kappa/4$. 
Thus, the conventional Pt/AFM interface response is intrinsically doubled in phase, while the shared-spacer term depends on the relative double phase $2(\phi_{\mathrm{T}}-\phi_{\mathrm{B}})$.

The oscillatory resistance dynamics described by \eqref{eq:smr-phase-resistance} provide a mechanism for extracting useful terahertz-frequency signal power on different harmonics of the investigated AFM dynamics.
For the difference-frequency analysis in Figure~\ref{fig:output-signatures}(E), we characterize the relative dynamics by a frequency shift $\Delta f= f_{\mathrm{B}}-f_{\mathrm{T}}$, where the signs distinguish the rotation directions of the two N\'eel vectors. 
This panel uses a separate 81-point sweep with $j_{\mathrm{T}}=2.0\times10^{8}$~A~cm$^{-2}$ fixed and $j_{\mathrm{B}}$ varied from 0 to $4.0\times10^{8}$~A~cm$^{-2}$ with increments of $0.05\times10^{8}$~A~cm$^{-2}$. 
Coupled and zero-coupling reference trajectories were solved independently over 60~ps, and their signed frequencies were extracted from the 30--60~ps interval. 
This protocol differs from the 100-ps principal simulations described below.

Figure~\ref{fig:device-circuit}(C,D) shows the necessary output AC signal extraction circuit. 
For a resistance component $R_{\mathrm{ac}}\cos(\omega_\star t+\psi)$ under a DC read (signal) current $I_0$, the SHO can be treated as a small-signal source with an output AC voltage $U_{\mathrm{ac}}=I_0R_{\mathrm{ac}}\cos(\omega_\star t+\psi)$. 
Here $R_\mathrm{ac}$ represents the Fourier component of the total DAL device magnetoresistance at frequency $f_\star = \omega_\star/2\pi$.
According to \eqref{eq:smr-phase-resistance}, $R_{\mathrm{ac}}$ is the amplitude of the resistance oscillations, $R_{\mathrm{T}}\cos(2\phi_{\mathrm{T}})
 + R_{\mathrm{B}}\cos(2\phi_{\mathrm{B}}) + R_{\mathrm{mix}}\cos[2(\phi_{\mathrm{T}}-\phi_{\mathrm{B}})]$, defined at $f_\star = \omega_\star/2\pi$. 
It is clear, that depending on the device operating regime, $R_{\mathrm{ac}}$ can be equal to $R_{\mathrm{T}}$, $R_{\mathrm{B}}$, $R_{\mathrm{mix}}$ or depend on their combinations.

To extract the generated AC power, the source should be connected in series to a passive resistance $R_0=\langle R\rangle$, which receives the DC current-driven DAL SHO's output (the similar technique was proposed in \cite{Sulymenko2018Tunnel}). 
Thus, the power transferred to a load resistor $R_{\mathrm{L}}$ is
\begin{equation}
 P_{\mathrm{L}}(\omega_{\star})=
 \frac{I_{0}^{2}R_{\mathrm{ac}}^{2}}{2}
 \frac{R_{\mathrm{L}}}{\left(R_{0}+R_{\mathrm{L}}\right)^{2}} \, .
 \label{eq:load-power}
\end{equation}
For matched loading, $R_{\mathrm{L}}=R_0$, so \eqref{eq:load-power} can be reduced to
\begin{equation}
 P_{\mathrm{L}}^{\mathrm{max}}(\omega_{\star})=
 \frac{I_{0}^{2}R_{\mathrm{ac}}^{2}}{8R_{0}} \, .
 \label{eq:matched-load-power}
\end{equation}

When a DAL spin Hall device is used to detect input terahertz-frequency signals, it generates an output signal in the form of a DC voltage $U_{\rm dc}$. 
This voltage is generated when an input signal current $I_0 \cos(\omega t)$ with an amplitude of $I_0$ and angular frequency of $\omega = \omega_\star$ is applied to the central Pt electrode of the structure. 
Due to the precise mixing of the $I_0 \cos(\omega t)$ and $R_{\mathrm{ac}}\cos(\omega_\star t+\psi)$ oscillations, a DC voltage component,
\begin{eqnarray}
 U_{\mathrm{dc}} &=&
 \frac{\omega_{\star}}{2\pi} \int_0^{2\pi/\omega_{\star}}I_{0} R_{\mathrm{ac}}\cos(\omega_\star t)\cos(\omega_\star t+\psi)
 \nonumber\\
 &=& \frac{I_0 R_{\rm ac}}{2} \cos(\psi)
 \, ,
 \label{eq:udc}
\end{eqnarray}
with a maximum value of $U^{\rm max}_{\mathrm{dc}} = I_0 R_{\rm ac}/2$ at $\psi = 0$, is generated.

Finally, Table~\ref{tab:parameters} summarizes the numerical inputs required to perform the subsequent simulations. 
Unless stated otherwise, the magnetic, material, geometrical, and transport parameters are adopted from the similar canted AFM and dual-free-layer oscillator studies \cite{Sulymenko2017Terahertz,Prokopenko2013Hysteresis,Shtanko2024DualAFM}. 
The single-interface SHMR ratio $a_{\mathrm{T,B}}/R_{\mathrm{b}}=2.5\times10^{-3}$ is based on the conventional SHMR contrast reported for hematite/Pt structures \cite{Chen2013SMR,Lebrun2019SMR}. 
In contrast, the shared-central-Pt mixing ratio $\kappa/R_{\mathrm{b}}=2.5\times10^{-4}$ is a phenomenological approximation introduced in this work; because an experimental value for this two-interface contribution is not available, it is conservatively chosen one order of magnitude below the single-interface SHMR ratio.

\begin{table}
\centering
\caption{Geometry, magnetic, transport, and numerical parameters used in the DAL spin Hall device simulations.}
\label{tab:parameters}
\renewcommand{\arraystretch}{1.08}
\begin{tabular}{@{}lll@{}}
\hline
Quantity & Symbol & Value \\
\hline
Outer Pt thickness & $d_{\mathrm{Pt}}$ & $20$~nm \\
AFM thickness & $d_{\mathrm{AFM}}$ & $5$~nm \\
Central Pt thickness & $l_{\mathrm{Pt}}$ & $12$~nm \\
Free layer radius & $r$ & $250$~nm \\
Readout length & $L$ & $500$~nm \\
Readout width & $w$ & $500$~nm \\
Exchange field & $H_{\mathrm{ex}}$ & $0.7\times10^9$~A~m$^{-1}$ \\
Hard-axis field & $H_{\mathrm{h}}$ & $15.9\times10^3$~A~m$^{-1}$ \\
Easy-plane field & $H_{\mathrm{e}}$ & $15.9$~A~m$^{-1}$ \\
DMI field & $H_{\mathrm{DMI}}$ & $1.75\times10^6$~A~m$^{-1}$ \\
Saturation magnetization & $M_s$ & $8.40\times10^5$~A~m$^{-1}$ \\
Intrinsic damping & $\alpha_0$ & $10^{-4}$ \\
Spin-mixing conductance & $g^{\uparrow\downarrow}$ & $6.9\times10^{18}$~m$^{-2}$ \\
Pt spin Hall angle & $\theta_{\mathrm{SH}}$ & $0.10$ \\
Pt spin-diffusion length & $\lambda_{\mathrm{Pt}}$ & $7.3$~nm \\
Pt resistivity & $\rho_{\mathrm{Pt}}$ & $4.8\times10^{-7}$~$\Omega$~m \\
Resistance of central Pt electrode & $R_0$ & $40$~$\Omega$ \\
Top-interface SHMR ratio & $a_{\mathrm{T}}/R_{\mathrm{b}}$ & $2.5\times10^{-3}$ \\
Bottom-interface SHMR ratio & $a_{\mathrm{B}}/R_{\mathrm{b}}$ & $2.5\times10^{-3}$ \\
Central-Pt mixing ratio & $\kappa/R_{\mathrm{b}}$ & $2.5\times10^{-4}$ \\
Simulation time interval & $t_{\mathrm{sim}}$ & $0$--$100$~ps \\
Spectral analysis interval & $t_{\mathrm{FFT}}$ & $50$--$100$~ps \\
\hline
\end{tabular}
\end{table}

\section{Results}\label{sec:results}

\begin{figure}
  \centering
  \includegraphics[width=\linewidth]{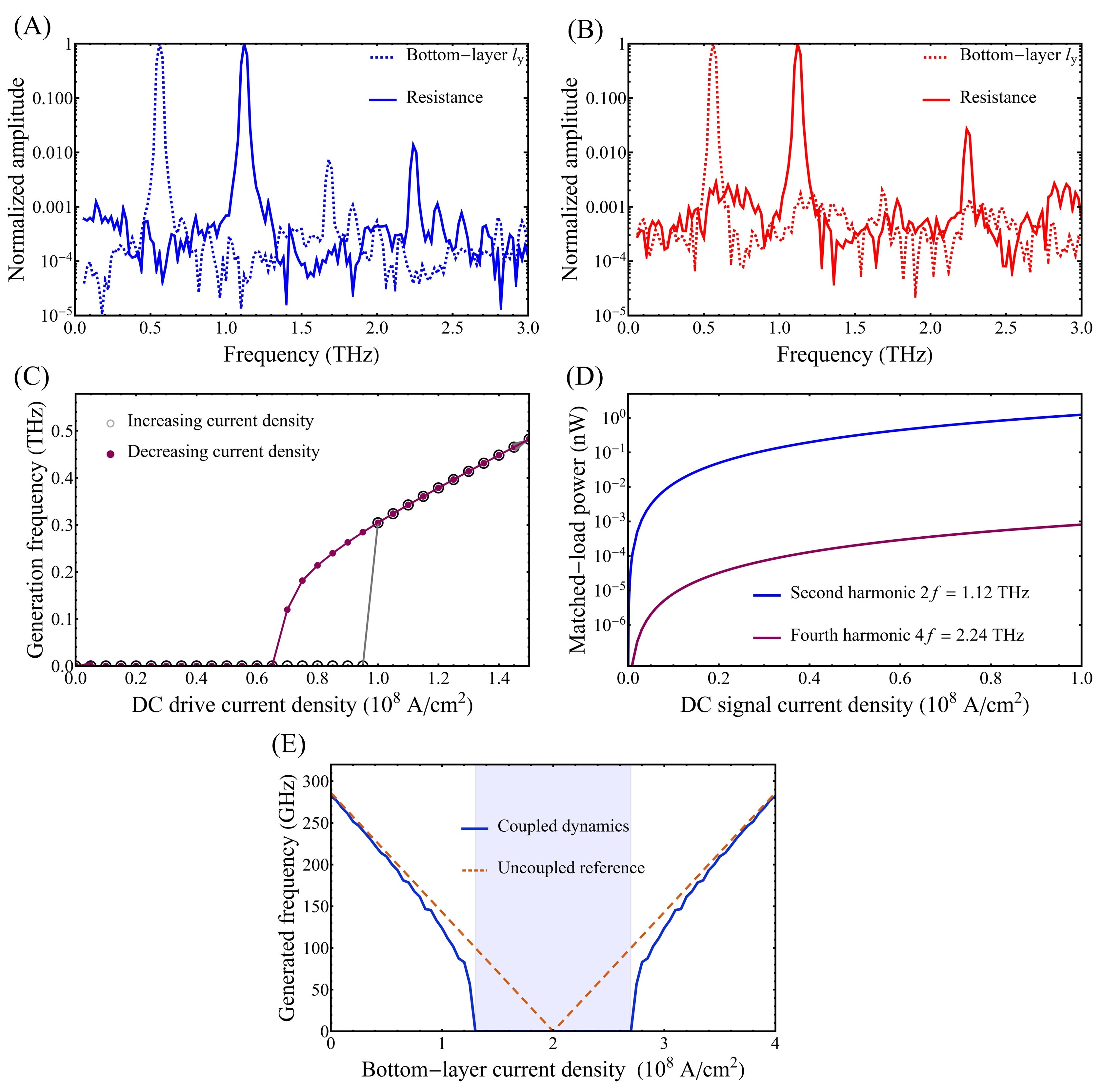}
  \caption{Electrical output signatures of the DAL SHO. 
  (A,B) Logarithmic one-sided spectra from the 50--100~ps steady-state interval for (A) co-rotation and (B) counter-rotation. Dashed curves are normalized bottom-layer $l_y^{(\mathrm{B})}$ spectra and solid curves are normalized AC resistance spectra. The N\'eel fundamental frequency is near $f=0.561$~THz, the conventional SHMR response is near $2f=1.12$~THz, and the counter-rotational phase-sensitive mixing contribution is near $4f=2.24$~THz. 
  (C) Hysteresis of the doubled generation frequency $2|f_{\mathrm{B}}|$ under increasing (open black-bordered) and decreasing (dark-magenta) current density. 
  (D) Matched-load power predicted for the counter-rotational second- and fourth-harmonic resistance components versus central-Pt read-current density. 
  (E) Generated difference frequency $|\Delta f| = |f_{\mathrm{B}}-f_{\mathrm{T}}|$ versus bottom-layer current density for fixed $j_{\mathrm{T}}=2.0\times10^{8}$~A~cm$^{-2}$. The solid blue curve shows the coupled dynamics, the dashed orange curve is the uncoupled reference, and the shaded region marks the co-rotational synchronization interval $1.3\leq j_{\mathrm{B}}/(10^{8}\,\mathrm{A\,cm^{-2}})\leq2.7$, where $\Delta f=0$.}
  \label{fig:output-signatures}
\end{figure}

To obtain numerical results, the LLGS system \eqref{eq:llgs} was integrated as a 12-component initial-value problem with the parameter values listed in Table~\ref{tab:parameters}. The initial sublattice states were antiparallel in both AFM layers,
\begin{eqnarray}
	\mathbf{m}_{1}^{(\mathrm{B})}(0)&=&\mathbf{m}_{1}^{(\mathrm{T})}(0)
	=(0.6,0,0.8)\,,
    \nonumber\\
	\mathbf{m}_{2}^{(\mathrm{B})}(0)&=&\mathbf{m}_{2}^{(\mathrm{T})}(0)
	=(-0.6,0,-0.8),
	\label{eq:initial-state}
\end{eqnarray}
which gives $\mathbf{l}^{(i)}(0)=(0.6,0,0.8)$ and $\boldsymbol{\mu}^{(i)}(0)=0$. 
The baseline drive current density magnitudes were taken to be $|j_{\mathrm{T}}|=3.4\times10^{8}$ and $j_{\mathrm{B}}=3.5\times10^{8}$~A~cm$^{-2}$. 
Adaptive stiffness-switching integration uses accuracy and precision goals of 9, unrestricted maximum step count, and 10,001 uniformly spaced samples over 0--100~ps. 
Spectra use the 50--100~ps interval after separate mean removal, a Hann window algorithm \cite{Harris1978Windows, Nuttall1981Windows}, and a one-sided discrete Fourier transform; the displayed native bin spacing is 0.02~THz. 
The continuation calculation in Figure~\ref{fig:output-signatures}(C) carries the terminal four-sublattice state between successive current steps.

The bottom-layer order parameter ${\bf l}^{(B)}$ moves along a stable, predominantly easy-plane orbit, as presented in Figure~\ref{fig:neel-trajectory}. 
The $l^{(B)}_y$--$l^{(B)}_z$ projection is nearly circular, whereas the $l^{(B)}_x$ projections reveal a smaller hard-axis excursion. 
The early ``non-closed'' part of the trajectory is due to the relaxation from the prescribed antiparallel state; the later recurrent orbit demonstrates sustained N\'eel-vector rotation. 
The unit-sphere representation in Figure~\ref{fig:neel-trajectory}(C) shows the weak non-circularity of the N\'eel-vector trajectory and the explicit out-of-plane canting. 
These deviations are expected from the simultaneous action of the anisotropy, DMI, and the interlayer dipolar field, and they provide a nonlinear source of higher resistance harmonics.

For the nearly detuned drive current magnitudes used in Figure~\ref{fig:output-signatures}(A,B), the late-time unwrapped phases give $|f_{\mathrm{T}}|\simeq|f_{\mathrm{B}}|\simeq0.561$~THz. 
The frequency signs are equal in the co-rotation and opposite in the counter-rotation regimes. 
Thus, the AFM layers are phase-locked despite the small current mismatch. 
A detailed analysis of synchronization transients, locking range, and phase dynamics is reported elsewhere \cite{Shtanko2026SynchronizationBandwidth}.

The dashed spectra in Figure~\ref{fig:output-signatures}(A,B) correspond to the bottom-layer N\'eel vector oscillations. 
The resistance spectra, shown by solid lines, have their largest AC peak near $f_{\mathrm{T}} + f_{\mathrm{B}} \approx 2f \approx 1.12$~THz because each conventional SHMR contribution contains $\cos(2\phi_i)$ (see (\eqref{eq:smr-phase-resistance})). The corresponding unnormalized one-sided amplitudes of $\Delta R/R_{\mathrm{dc}}$ are $2.56\times10^{-3}$ for co-rotation and $2.58\times10^{-3}$ for counter-rotation.

In co-rotation, an ideal circular orbit would make the mixing phase $\phi_{\mathrm{T}}-\phi_{\mathrm{B}}$ static. 
The observed feature near $4f\simeq2.24$~THz is consequently dominated by the non-circular order parameter orbit and the resulting non-sinusoidal interface projections: the top and bottom interface terms contribute $1.68\times10^{-5}$ and $1.67\times10^{-5}$, whereas the mixing term contributes $1.65\times10^{-6}$. 
Their combined fourth-harmonic amplitude is $3.47\times10^{-5}$.

In counter-rotation, $\phi_{\mathrm{T}}\simeq\omega t+\phi_{\mathrm{T},0}$ and $\phi_{\mathrm{B}}\simeq-\omega t+\phi_{\mathrm{B},0}$, so the central-Pt term in Eq.~\eqref{eq:smr-phase-resistance}, $R_{\mathrm{mix}}\cos[2(\phi_{\mathrm{T}}-\phi_{\mathrm{B}})]$, varies directly with an angular frequency of $4\omega$. 
At this harmonic, the mixing contribution rises to $5.91\times10^{-5}$, while the top and bottom interface terms are $4.92\times10^{-6}$ and $4.98\times10^{-6}$; the total amplitude is therefore $6.89\times10^{-5}$, a factor of $1.98$ larger than in co-rotation.
Because panels (A) and (B) are independently peak-normalized, this factor is determined from the unnormalized spectral data, not from their displayed heights. 
The small broadband features outside the resolved harmonics arise from the nonlinear LLGS dynamics in Eq.~\eqref{eq:llgs} and the finite analysis window; they should not be interpreted as additional fundamental modes.

The continuation analysis corresponding to the doubled output frequency, $2\vert{}f_{\mathrm{B}}\vert{}$, is illustrated in Figure~\ref{fig:output-signatures}(C). 
On the increasing-current branch, the static state persists up to $j_B\simeq1.0\times10^{8}$~A~cm$^{-2}$, where the rotating solution is reached. 
Once generated, the rotating state remains stable on the decreasing-current branch down to $j_B\simeq0.7\times10^{8}$~A~cm$^{-2}$. 
Thus, the numerically resolved hysteresis interval is approximately $0.3\times10^{8}$~A~cm$^{-2}$; its endpoints are limited by the $0.05\times10^{8}$~A~cm$^{-2}$ continuation increment. 
Its physical origin in the present LLGS model is the corresponding-sublattice mutual dipolar field: it makes the state of either AFM layers depend on the state of the other. 
When the device is approached from rest, the quasistatic dipolar field helps preserve the metastable static configuration until the upper threshold is exceeded. 
Once the two N\'eel vectors rotate, their phase-locked, time-dependent dipolar fields self-consistently stabilize the rotating branch, allowing it to persist down to the lower threshold. 
The fact that the nucleation current exceeds the sustaining current is therefore a dipolar-coupling-induced bistability, analogous to the hysteretic counter-precession reported for dual-free-layer spin-torque nano-oscillators \cite{Prokopenko2013Hysteresis}, rather than a plotting or fitting artifact. 

Figure~\ref{fig:output-signatures}(D) translates the counter-rotational resistance components into matched-load power. 
At a fixed resistance amplitude, Eq.~\eqref{eq:matched-load-power} gives $P\propto I_0^2$; the two curves consequently increase quadratically with read-current density. 
At the largest modeled read-current density, $j_{\mathrm{S}}=1.0\times10^{8}$~A~cm$^{-2}$ ($I_0=6.0$~mA), the maximum matched-load powers are $1.236$~nW at the double frequency of $2f$ and $0.807$~pW at the quadruple frequency of $4f$. 
The second-harmonic power exceeds the fourth-harmonic power because the resistance amplitude at $2f$ is much larger, even though the counter-rotational mixing mechanism enhances the relative resistance response at $4f$.

The difference-frequency channel, facilitated by the relative-phase sensitivity of $R_{\mathrm{mix}}$ defined in \eqref{eq:smr-phase-resistance}, is presented in Figure~\ref{fig:output-signatures}(E). 
Over the co-rotational locking interval, $1.3\leq j_{\mathrm{B}}/(10^{8}\,\mathrm{A\,cm^{-2}})\leq2.7$, the coupled top- and bottom-layer frequencies coincide and $\Delta f$ collapses to zero. 
Outside this interval, the coupled curve recovers a nonzero output that is generally of order $10^{2}$~GHz; the smallest resolved nonzero value immediately adjacent to the locking boundaries is approximately 56.5~GHz. 
The uncoupled reference instead approaches zero smoothly at $j_{\mathrm{B}}=j_{\mathrm{T}}$. 
This isolated natural-frequency crossing does not constitute synchronization because no interaction maintains frequency equality over a finite current interval.

\begin{figure}
  \centering
  \includegraphics[width=\linewidth]{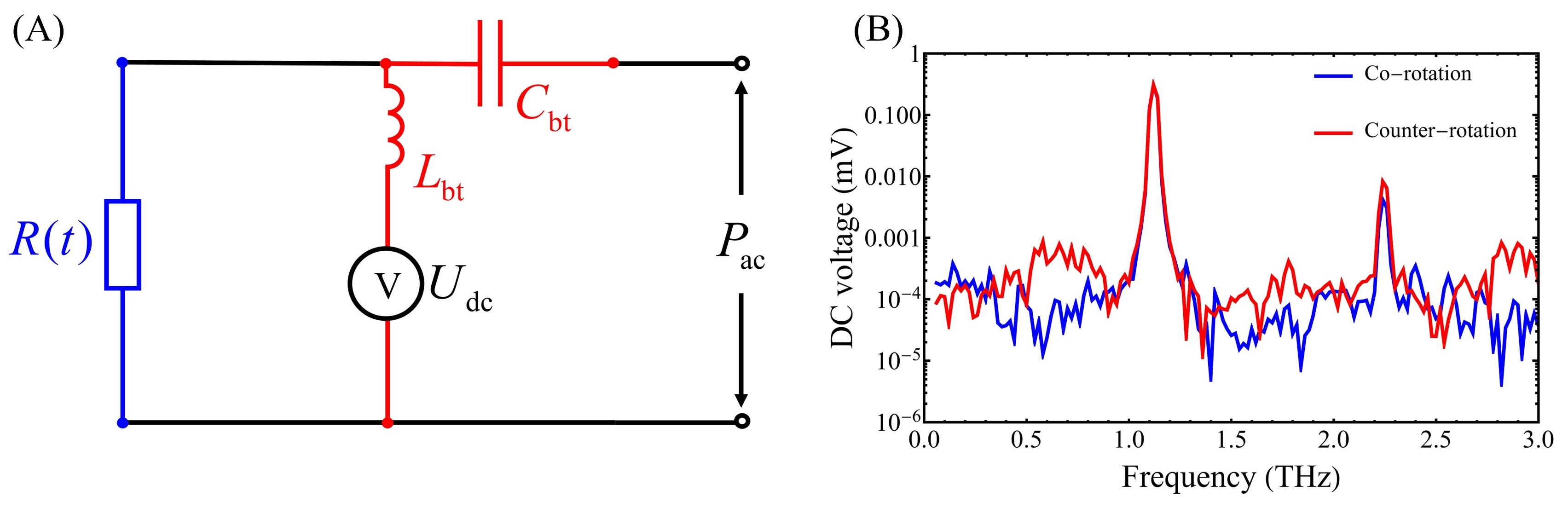}
  \caption{The DAL SHD equivalent scheme and its frequency-dependent output. 
  (A) The small-signal Thevenin representation of the detector as an oscillating resistance $R(t)$ connected via an ideal bias tee to an external signal source with power $P_{\rm ac}$.
  (B) The maximum output DC voltage $U^{\rm max}_{\rm dc}$ generated in a DAL detector as a function of input signal frequency $f_\star$.   The detector operates in a regime where the input signal frequency and the device's magnetoresistance frequency are perfectly matched.
  }
  \label{fig:detector}
\end{figure}

Figure~\ref{fig:detector}(A) shows an equivalent scheme of a DAL SHD connected via an ideal bias tee to a source of an external terahertz-frequency signal.
This scheme primarily resembles previous schemes of other AFM detectors \cite{Artemchuk2020ATJ,Prokopenko2024Book}, and highlights the principle of converting the input signal current with power $P_{\rm ac}$ into an output DC voltage $U_{\rm dc}$.
According to Eq.~(\eqref{eq:udc}), this voltage is proportional to the AC resistance amplitude $R_{\rm ac}$ at the input signal frequency $f = f_\star$. 
Figure~\ref{fig:detector}(B) presents the frequency dependence of the maximum DC voltage, $U^{\rm max}_{\rm dc}$, calculated from the data shown in Figure~\ref{fig:output-signatures}(A,B) using equation (\eqref{eq:udc}).

\section{Discussion}\label{sec:discussion}

The DAL device architecture separates excitation and readout functions: the outer Pt layers control the sign of the spin Hall torque in each AFM layers, whereas the central Pt layer senses both N\'eel vectors. 
The predominantly easy-plane but weakly non-circular orbit in Figure~\ref{fig:neel-trajectory} provides the magnetic origin of the higher-harmonic content. 
This separation permits two qualitatively distinct phase relations without changing the local frequency scale. 
The ordinary SHMR terms yield a robust frequency-doubled electrical signal in either rotational regime, as shown by the $2f$ peaks in Figure~\ref{fig:output-signatures}(A,B). 
The central Pt layer is not merely two independent interface resistors: the shared spin accumulation is represented by the phenomenological mixing term proportional to $\kappa$, which depends on the relative orientations of the two N\'eel vectors. 
In co-rotation this relative phase is nearly stationary, so the mixing contribution does not supply an ideal circular-orbit AC signal. 
In counter-rotation, it evolves at twice the magnetic angular frequency, and the central-Pt mixing term therefore appears at the fourth harmonic, accounting for the enhanced $4f$ feature in Figure~\ref{fig:output-signatures}(B). 
The approximately twofold enhancement found here is not a generic universal factor; it follows from the present N\'eel vector orbit and the adopted ratio $\kappa/a_{\mathrm{T,B}}=0.1$.

The finite zero-difference interval in Figure~\ref{fig:output-signatures}(E) shows that the same coupling that stabilizes collective dynamics also limits low-frequency steady-state operation. 
Rather than yielding an arbitrarily small nonzero $\Delta f$ as the uncoupled frequencies approach one another, the interacting oscillators enter a synchronized state and the difference-frequency channel becomes static. 
This behavior is consistent with the independently studied finite synchronization band and its reduction as the central-Pt spacer is made thicker \cite{Shtanko2026SynchronizationBandwidth}.

Accordingly, two prospective ways could extend operation toward lower nonzero frequencies. 
First, drive-current pulses shorter than the synchronization build-up time could access the transient detuned response before phase locking is established. 
Second, a thicker central read-Pt layer could weaken the interlayer coupling and narrow the locking interval. 
The latter approach involves a readout trade-off because increasing the spacer thickness may also reduce the shared magnetoresistive contribution and hence the measurable output signal amplitude. 
Neither strategy is simulated here; both require dedicated transient-dynamics and spin-transport calculations.

The continuation branches in Figure~\ref{fig:output-signatures}(C) demonstrate that a useful output state is history dependent near threshold. 
This is relevant for current-controlled operation because the accessible generation state depends on whether the device is approached from the static or rotating branch. 

The power curves in Figure~\ref{fig:output-signatures}(D) should be interpreted as small-signal matched-load benchmarks, not as an optimized radiated-power prediction. 
A practical route to higher generated power is an array of nominally matched DAL SHOs with their central-Pt readout voltages combined in a low-loss, impedance-matched network. 
If the selected $2f$ or $4f$ outputs are phase coherent, the AC voltages add linearly with the number of devices $N$, so the power delivered to a fixed matched load can, in principle, scale as $N^2$. 
In contrast, uncorrelated devices provide only incoherent power addition, which scales as $N$. 
Realizing the coherent regime requires control of frequency dispersion, phase locking, interconnect loss, and load matching; it is therefore a device- and array-level design problem beyond the present single-DAL-SHO calculation. 
The power estimates assume a 40~$\Omega$ central-Pt resistance, a non-invasive read current, and read-current-independent SHMR amplitudes. 
Likewise, the magnetic calculation is a zero-temperature macrospin model with phenomenological shared-spacer mixing; it omits thermal phase noise, nonuniform modes, detailed spin diffusion, and read-current torque. 
These limitations set clear priorities for device-scale micromagnetics and transport experiments. 
Although the present calculations use synchronization only to delimit the accessible difference-frequency range, its mechanism, transient formation, and parameter dependence are analyzed in detail in \cite{Shtanko2026SynchronizationBandwidth}.

Figure~\ref{fig:detector}(B) shows that, in the ``ideal mixing'' regime, when the input signal frequency fully coincides with the device's AC magnetoresistance frequency $f_{\star}$, the detector output simply mirrors the frequency dependence of magnetoresistance $R_{\rm ac}(f)$, which varies substantially in magnitude across a wide frequency range (see Figure~\ref{fig:output-signatures}(A,B)). 
Note that in the co- and counter-rotation regimes, the output DC voltages reach $U_{\rm dc} \approx 400$~$\mu$V at frequency $f_\star = 2f$, and $U_{\rm dc} \approx 10$~$\mu$V at the frequency $f_\star = 4f$, which is sufficient for some applications. 
At frequencies far from $2f$ and $4f$, the generated voltage is lower than 1~$\mu$V. 
This demonstrates the important role of coupled collective dynamics in both AFM layers of the DAL structure for its efficient operation as a signal detector.

\section{Conclusions}\label{sec:conclusions}

We developed a Pt/AFM/Pt/AFM/Pt spin Hall device model with two canted AFM layer, independent outer-Pt layer excitation and a central Pt layer magnetoresistance readout. 
The canted antiferromagnetic dynamics support locked co- and counter-rotational N\'eel-vector states near the frequency of $f \approx 0.561$~THz. 
These states generate a doubled electrical response near $2f \approx 1.12$~THz, while counter-rotation activates the phase-sensitive central-Pt contribution at $4f \approx 2.24$~THz and yields enhancement of the fourth harmonic by a factor of two. 
During a co-rotational current sweep, the shared-central-Pt response provides a difference-frequency channel of order $10^{2}$~GHz, but synchronization collapses this output to zero over a finite locking interval and prevents continuous steady-state tuning toward arbitrarily low nonzero frequencies. 
Pulsed operation shorter than the locking transient or a thicker central-Pt spacer may extend the accessible range, though the latter can reduce the magnetoresistive readout. 
The considered device can produce an output AC signal power of $1.236$~nW at a frequency of $2f$ and $0.807$~pW at a frequency of $4f$ if DC current of $6.0$~mA is applied to the central Pt electrode. 
When the applied current is a harmonic signal with a frequency equal to that of the device's magnetoresistance oscillations, the device operates as a signal detector and produces an output DC voltage of approximately 400~$\mu$V at $2f$ and about 10~$\mu$V at $4f$. 
The magnitudes of the output signals obtained in the oscillation and detection regimes of the device are sufficient for some applications. 
We believe that the dual-antiferromagnetic-layer spin Hall geometry is a promising approach for developing tunable, multi-frequency signal sources and detectors that operate in the microwave and terahertz bands.

\section*{Funding}
This work was supported by grant No.~2025.07/0237 from the National Research Foundation of Ukraine and partially by the Science and Technology Center in Ukraine (2025 IEEE Magnetism in Ukraine Initiative) under grant No. 9918.

\section*{Acknowledgments}
The authors would like to thank all the brave defenders of Ukraine who made the finalization of this publication possible.


\begin{thebibliography}{99}

\bibitem{Chen2024Emerging}
Chen H, Liu L, Zhou X, Meng Z, Wang X, Duan Z, et al. Emerging antiferromagnets for spintronics. Adv Mater (2024) 36:2310379. \href{https://doi.org/10.1002/adma.202310379}{doi: 10.1002/adma.202310379}


\bibitem{Rimmler2025Noncollinear}
Rimmler BH, Pal B, Parkin SSP. Non-collinear antiferromagnetic spintronics. Nat Rev Mater (2025) 10:109--127. \href{https://doi.org/10.1038/s41578-024-00706-w}{doi: 10.1038/s41578-024-00706-w}


\bibitem{Sakamoto2025Diode}
Sakamoto S, Nomoto T, Higo T, Hibino Y, Yamamoto T, Tamaru S, et al. Antiferromagnetic spin-torque diode effect in a kagome Weyl semimetal. Nat Nanotechnol (2025) 20:216--221. \href{https://doi.org/10.1038/s41565-024-01820-0}{doi: 10.1038/s41565-024-01820-0}


\bibitem{Yao2025Transistors}
Yao F, Liao M, Gibertini M, Cheon CY, Lin X, Wu F, et al. Switching on and off the spin polarization of the conduction band in antiferromagnetic bilayer transistors. Nat Nanotechnol (2025) 20:609--616. \href{https://doi.org/10.1038/s41565-025-01872-w}{doi: 10.1038/s41565-025-01872-w}


\bibitem{Cheng2025Antiferromagnet}
Cheng Y, Zhou F, Teng J, Li P, Jiang L, Gong P, et al. Antiferromagnet-topological insulator heterostructure for polarization-controllable terahertz generation. Nat Commun (2025) 16:5656. \href{https://doi.org/10.1038/s41467-025-60060-5}{doi: 10.1038/s41467-025-60060-5}


\bibitem{Rene2025Terahertz}
Ren{\'e} S, Levchuk A, Abdelsamie A, Li Z, Dufour P, Chaudron A, et al. Terahertz antiferromagnetic dynamics induced by ultrafast spin currents. Sci Adv (2025) 11:eadx1107. \href{https://doi.org/10.1126/sciadv.adx1107}{doi: 10.1126/sciadv.adx1107}


\bibitem{Ogawa2026Ultrafast}
Ogawa K, Tsai H, Yoshikawa N, Matsuo T, Tsushima Y, Asakura M, et al. Ultrafast time-resolved observation of non-thermal current-induced switching in an antiferromagnetic Weyl semimetal. Nat Mater (2026) 25:434--439. \href{https://doi.org/10.1038/s41563-025-02402-8}{doi: 10.1038/s41563-025-02402-8}


\bibitem{Lee2025SpinTorque}
Lee WB, Hwang S, Ko HW, Park BG, Lee KJ, Choi GM. Spin-torque-driven gigahertz magnetization dynamics in the non-collinear antiferromagnet Mn$_{3}$Sn. Nat Nanotechnol (2025) 20:487--493. \href{https://doi.org/10.1038/s41565-025-01859-7}{doi: 10.1038/s41565-025-01859-7}


\bibitem{Cheng2016Terahertz}
Cheng R, Xiao D, Brataas A. Terahertz antiferromagnetic spin Hall nano-oscillator. Phys Rev Lett (2016) 116:207603. \href{https://doi.org/10.1103/PhysRevLett.116.207603}{doi: 10.1103/PhysRevLett.116.207603}


\bibitem{Khymyn2017Josephson}
Khymyn R, Lisenkov I, Tiberkevich V, Ivanov BA, Slavin A. Antiferromagnetic THz-frequency Josephson-like oscillator driven by spin current. Sci Rep (2017) 7:43705. \href{https://doi.org/10.1038/srep43705}{doi: 10.1038/srep43705}


\bibitem{Gomonay2010SpinTransfer}
Gomonay HV, Loktev VM. Spin transfer and current-induced switching in antiferromagnets. Phys Rev B (2010) 81:144427. \href{https://doi.org/10.1103/PhysRevB.81.144427}{doi: 10.1103/PhysRevB.81.144427}


\bibitem{Fischer2018SpinHall}
Fischer J, Gomonay O, Schlitz R, Ganzhorn K, Vlietstra N, Althammer M, et al. Spin Hall magnetoresistance in antiferromagnet/heavy-metal heterostructures. Phys Rev B (2018) 97:014417. \href{https://doi.org/10.1103/PhysRevB.97.014417}{doi: 10.1103/PhysRevB.97.014417}


\bibitem{Sinova2015SpinHall}
Sinova J, Valenzuela SO, Wunderlich J, Back CH, Jungwirth T. Spin Hall effects. Rev Mod Phys (2015) 87:1213--1260. \href{https://doi.org/10.1103/RevModPhys.87.1213}{doi: 10.1103/RevModPhys.87.1213}


\bibitem{Miyamoto2024WideBand}
Miyamoto M, Kobayashi R, Kuwano G, Tsujimoto M, Kakeya I. Wide-band frequency modulation of a terahertz intrinsic Josephson junction emitter of a cuprate superconductor. Nat Photonics (2024) 18:267--275. \href{https://doi.org/10.1038/s41566-023-01348-0}{doi: 10.1038/s41566-023-01348-0}


\bibitem{Karimi2024Bolometric}
Karimi B, Steffensen GO, Higginbotham AP, Marcus CM, Levy Yeyati A, Pekola JP. Bolometric detection of Josephson radiation. Nat Nanotechnol (2024) 19:1613--1618. \href{https://doi.org/10.1038/s41565-024-01770-7}{doi: 10.1038/s41565-024-01770-7}


\bibitem{Khymyn2018Neuron}
R. Khymyn, I. Lisenkov, J. Voorheis \emph{et al.},
``Ultra-fast artificial neuron: generation of picosecond-duration spikes in a current-driven antiferromagnetic auto-oscillator,''
{\emph{Sci. Rep.}, Vol. 8, p. 15727, Oct. 2018.}


\bibitem{Sulymenko2018Logic}
O. Sulymenko, O. Prokopenko, I. Lisenkov \emph{et al.},
``Ultra-fast logic devices using artificial ``neurons'' based on antiferromagnetic pulse generators,''
{\emph{J. Appl. Phys.}, Vol. 124, p. 152115, Sep. 2018.}


\bibitem{Sotnyk2026PNeuron}
I. Sotnyk, O. Prokopenko, 
“Antiferromagnetic Programmable Neuron: Structure, Training, and Pattern Recognition Applications,” 
IEEE Journal on Exploratory Solid-State Computational Devices and Circuits 12, 9–17 (2026). 


\bibitem{Sulymenko2017Terahertz}
Sulymenko OR, Prokopenko OV, Tiberkevich VS, Slavin AN, Ivanov BA, Khymyn RS. Terahertz-frequency spin Hall auto-oscillator based on a canted antiferromagnet. Phys Rev Appl (2017) 8:064007. \href{https://doi.org/10.1103/PhysRevApplied.8.064007}{doi: 10.1103/PhysRevApplied.8.064007}


\bibitem{Sulymenko2018Tunnel}
Sulymenko OR, Prokopenko OV, Tyberkevych VS, Slavin AN. Terahertz-frequency signal source based on an antiferromagnetic tunnel junction. IEEE Magn Lett (2018) 9:3104605. \href{https://doi.org/10.1109/LMAG.2018.2852291}{doi: 10.1109/LMAG.2018.2852291}


\bibitem{Park2011SpinValve}
Park BG, Wunderlich J, Mart{\'i} X, Hol{\'y} V, Kurosaki Y, Yamada M, et al. A spin-valve-like magnetoresistance of an antiferromagnet-based tunnel junction. Nat Mater (2011) 10:347--351. \href{https://doi.org/10.1038/nmat2983}{doi: 10.1038/nmat2983}


\bibitem{Xu2025SpinFlop}
Xu S, Zhang Z, Mahfouzi F, Huang Y, Cheng H, Dai B, et al. Giant spin-flop magnetoresistance in a collinear antiferromagnetic tunnel junction. Nat Commun (2025) 16:8370. \href{https://doi.org/10.1038/s41467-025-62695-w}{doi: 10.1038/s41467-025-62695-w}


\bibitem{Khymyn2017SHD}
R.~Khymyn, V.~Tiberkevich, and A.~Slavin,
``Antiferromagnetic spin current rectifier,''
{\emph{AIP Adv.}, Vol. 7, p. 055931, Mar. 2017.}


\bibitem{Prokopenko2024Book}
V. Prokopenko and O. Prokopenko, “Terahertz signal detectors based on antiferromagnetic spintronic nanostructures,” Chapter 4 in book: “Nanocomposite and Nanocrystalline Materials and Coatings – Microstructure, Properties and Applications,” Advanced Structured Materials (Vol. 214, Eds. A.D. Pogrebnjak, Y. Bing, M. Sahul), Singapore: Springer, 2024, P. 129-147. 


\bibitem{Artemchuk2020ATJ}
P.~Yu.~Artemchuk, O.~R.~Sulymenko, S.~Louis \emph{et al.},
``Terahertz frequency spectrum analysis with a nanoscale antiferromagnetic tunnel junction,''
{\emph{J. Appl. Phys.}, Vol. 127, p. 063905, Feb. 2020.}


\bibitem{Prokopenko2013Hysteresis}
Prokopenko OV, Krivorotov IN, Bankowski EN, Meitzler TJ, Tiberkevich VS, Slavin AN. Hysteresis regime in the operation of a dual-free-layer spin-torque nano-oscillator with out-of-plane counter-precessing magnetic moments. J Appl Phys (2013) 114:173904. \href{https://doi.org/10.1063/1.4828866}{doi: 10.1063/1.4828866}


\bibitem{Shtanko2024DualAFM}
Shtanko O, Prokopenko OV. Terahertz-frequency signals generated in a dual-antiferromagnetic-layer spin Hall oscillator. In: 2024 IEEE 14th International Conference Nanomaterials: Applications \& Properties (NAP). IEEE (2024) pp. 1--4. \href{https://doi.org/10.1109/NAP62956.2024.10739685}{doi: 10.1109/NAP62956.2024.10739685}


\bibitem{Zhao2025InterlayerDMI}
Zhao X, Zhao J. Interlayer Dzyaloshinskii-Moriya interaction in spintronics. npj Spintronics (2025) 3:25. \href{https://doi.org/10.1038/s44306-025-00091-2}{doi: 10.1038/s44306-025-00091-2}


\bibitem{Yang2023FirstPrinciplesDMI}
Yang H, Liang J, Cui Q. First-principles calculations for Dzyaloshinskii--Moriya interaction. Nat Rev Phys (2023) 5:43--61. \href{https://doi.org/10.1038/s42254-022-00529-0}{doi: 10.1038/s42254-022-00529-0}


\bibitem{Kim2016Ultrafast}
Kim TH, Gr{\"u}nberg P, Han SH, Cho B. Ultrafast spin dynamics and switching via spin transfer torque in antiferromagnets with weak ferromagnetism. Sci Rep (2016) 6:35077. \href{https://doi.org/10.1038/srep35077}{doi: 10.1038/srep35077}


\bibitem{Cheng2014SpinPumping}
Cheng R, Xiao J, Niu Q, Brataas A. Spin pumping and spin-transfer torques in antiferromagnets. Phys Rev Lett (2014) 113:057601. \href{https://doi.org/10.1103/PhysRevLett.113.057601}{doi: 10.1103/PhysRevLett.113.057601}


\bibitem{Hattori2024SpinPumping}
Hattori T, Fukuoka A, Kawagita K, Tada K, Ishikawa Y, Yamaguchi Y, et al. Inter- and intrasublattice spin mixing conductance of the antiferromagnetic spin pumping effect in $\alpha$-Fe$_2$O$_3$/Pt. Phys Rev Lett (2024) 133:256701. \href{https://doi.org/10.1103/PhysRevLett.133.256701}{doi: 10.1103/PhysRevLett.133.256701}


\bibitem{Dmytriiev2010CrossTensor}
O. Dmytriiev, T. Meitzler, E. Bankowski, A. Slavin, and V. Tiberkevich,
J. Phys.: Condens. Matter 22, 136001 (2010).

\bibitem{Chen2013SMR}
Chen YT, Takahashi S, Nakayama H, Althammer M, Goennenwein STB, Saitoh E, et al. Theory of spin Hall magnetoresistance. Phys Rev B (2013) 87:144411. \href{https://doi.org/10.1103/PhysRevB.87.144411}{doi: 10.1103/PhysRevB.87.144411}


\bibitem{Lebrun2019SMR}
Lebrun R, Ross A, Gomonay O, Bender SA, Baldrati L, Kronast F, et al. Anisotropies and magnetic phase transitions in insulating antiferromagnets determined by a spin-Hall magnetoresistance probe. Commun Phys (2019) 2:50. \href{https://doi.org/10.1038/s42005-019-0150-8}{doi: 10.1038/s42005-019-0150-8}


\bibitem{Harris1978Windows}
Harris FJ. On the use of windows for harmonic analysis with the discrete Fourier transform. Proc IEEE (1978) 66:51--83. \href{https://doi.org/10.1109/PROC.1978.10837}{doi: 10.1109/PROC.1978.10837}


\bibitem{Nuttall1981Windows}
Nuttall AH. Some windows with very good sidelobe behavior. IEEE Trans Acoust Speech Signal Process (1981) 29:84--91. \href{https://doi.org/10.1109/TASSP.1981.1163506}{doi: 10.1109/TASSP.1981.1163506}


\bibitem{Shtanko2026SynchronizationBandwidth}
Shtanko O, Prokopenko O. Synchronization bandwidth engineering in a dual-antiferromagnetic-layer spin Hall oscillator. In: Proceedings of the 2026 IEEE 7th KhPI Week on Advanced Technology (KhPIWeek); 5--9 October 2026; Kharkiv, Ukraine. IEEE (2026). In press.

\end{thebibliography}
\end{document}